\documentstyle[preprint,aps,epsf,psfig,floats]{revtex}
\begin{document}
\draft


\title{Phase Transitions in Nonequilibrium Systems}

\author{David Mukamel}

\address{Department of Physics of Complex Systems, The Weizmann Institute 
         of Science, Rehovot 76100, Israel\\
        [-2mm] $ $}

\maketitle
%
%

\vspace{2mm}
%
%
%

\section{I\lowercase{ntroduction}}

Collective phenomena in systems far from thermal equilibrium have been
a subject of extensive studies in recent years. Usually these systems
are driven out of equilibrium by external fields, such as electric
field in the case of conductors, pressure gradient in the case of
fluids, temperature gradient in the case of heat conductors, chemical
potential gradient in the case of growth problems and many others
\cite{Zia,KrugSpohn,Halpin,KrugAdvances}.  These driving fields are
very common in nature and are found in a large variety of physical
systems such as granular\index{granular flow} and traffic
flow\index{traffic flow} \cite{Wolf,Nagel1,Nagel2}, gel
electrophoresis\index{gel electrophoresis} \cite{Gel1,Gel2}, superionic
conductors\index{superioinc conductors} \cite{Superionic1,Superionic2}
to give a few examples.  In many cases these systems reach a steady
state, which unlike the equilibrium case, is characterised by
non-vanishing currents. In these lectures we consider possible collective
phenomena and phase transitions which may take place in such steady
states.  The main problem in studying nonequilibrium systems is the
lack of general theoretical framework within which they could be
analysed. As a result they are far less understood as compared with
equilibrium systems where the Gibbs picture provides such a theoretical
framework.

Before discussing nonequilibrium systems it is useful to consider
briefly systems in {\it thermal equilibrium}. Here decades of studies
have yielded a fairly detailed understanding of their thermodynamic
behaviour.  Many rules which govern phase transitions occurring in
these systems have been derived. For example, it has been shown that
the critical exponents associated with a phase transition may be
classified into {\it universality classes}\index{universality
classes}.  These classes do not depend on the detailed interactions in
the system but rather on a few parameters such as the symmetry of the
system and of the order parameter associated with the transition, the
dimensionality of the system and the range of interactions. Therefore,
in order to study theoretically the critical behaviour of a given
system it is sufficient to analyse the simplest possible model which
belongs to the same universality class. For reviews see, for example,
\cite{Wilson,Fisher}. It has also been shown that phase transitions
and spontaneous symmetry breaking do not take place at low
dimension. In particular, no phase transition is expected to take
place in a one dimensional system at finite temperatures as long as
the interactions are short range \cite{Landau}.  Moreover, breaking of
continuous symmetry may take place under the same conditions only in
dimensions higher than two \cite{Mermin}.  Other rules derived by
Landau relate the nature of the transition, namely whether it is first
order or continuous, to the symmetry of the systems \cite{Landau}.  If
the symmetry allows a third order term in the expansion of the free
energy in the order parameter, such as in the case of the transition
from a liquid to a nematic liquid crystal phase \cite{deGennes}, the
transition cannot be continuous and is necessarily first order. On the
other hand if the symmetry is such that no third order term is
allowed, such as in the transition from a paramagnetic to a
ferromagnetic phase, the transition may either be first order or
continuous, depending on the details of the interactions.  The Gibbs
phase rule\index{Gibbs phase rule} is another very useful example of a
rule which governs the phase digrams of systems in equilibrium
\cite{Landau}.  It deals with fluids composed of $c$ components. The
thermodynamic phase space of such systems is of $c+1$ dimensions,
associated with the temperature, pressure, and $c-1$ chemical
potentials. According to the rule, the manifold in this space on which
$n$ different phases coexist is of $D=2+c-n$ dimensions.  Another rule
deals with the phase diagram near a triple point, where three
coexistence lined meet. According to this $180^\circ$ rule, each of
the three angles defined by the intersecting coexistence lines must be
less than $180^\circ$.  This is a direct result of the convexity of
the free energy \cite{Wheeler}.

These rules and many others, some of which  are related to 
disordered systems\index{disordered systems} \cite{Harris,IM},
provide extremely useful tools for analysing and understanding phase diagrams
and critical behaviour of models and physical systems in equilibrium.
By simply identifying the symmetry of the system and the nature of the 
order paramenter involved in the phase transition one can usually
find the universality class of the transition and even obtain a rough idea of
the possible phase diagram.

Our degree of understanding of collective behaviour far from thermal
equilibrium is at a much more primitive stage. Since a general theoretical
framework for studying nonequilibrium phenomena does not exist, one cannot
derive similar rules which would be as general as those for equilibrium
systems. Rather, one has to resort to studying specific models and probe 
the resulting types of phase diagrams and phase transitions, with the hope
that some general picture might emerge.

In the present lectures we consider stochastic driven systems in one dimension
and discuss some interesting collective behaviour which they
display. Unlike equilibrium one dimensional systems which do 
not exhibit phase transitions, non equilibrium systems exhibit
a rich variety of collective phenomena such as first order and
continuous phase transitions, spontaneous symmetry breaking (SSB),
phase separation, slow coarsening processes and many others.
Mechanisms which lead to these phenomena are discussed.

The article is organised as follows: in Section~\ref{DB} the concept of
detailed balance is discussed, the lack of which is characteristic of
nonequilibrium systems. A necessary and sufficient condition for the
existence of detailed balance is presented. In Section~\ref{AEP} a
simple driven model, the totally asymmetric exclusion process, is
introduced and its phase diagram for a system with open boundaries is
calculated using a mean field approximation.  The phase diagram
exhibits several phases separated by first order and continuous
transitions. The matrix method which enables one to obtain exact
results for steady state properties is outlined in
Section~\ref{Matrix}.  A model which displays spontaneous symmetry
breaking in one dimension is introduced in Section~\ref{SSB} and a
model exhibiting phase separation accompanied by slow coarsening
processes is described in Section~\ref{PhaseSeparation}.  Open
problems and perspectives are briefly discussed in
Section~\ref{Summary}.

\section{D\lowercase{etailed} \lowercase{balance} \lowercase{and}
\lowercase{driven systems}}
\index{detailed balance} \label{DB}

In this section we make some general considerations concerning the 
evolution of dynamical systems. Let $C$ be a microscopic configuration,
and let $P(C,t)$ be the probability that the system
is in the microscopic configuration $C$ at time $t$. The dynamics
of the system is defined in terms of the transition rates
$W(C \to C')$ from a configuration $C$ to $C'$.  The equation
which governs the evolution of the distribution function $P(C,t)$
takes the form
\begin{equation}
\label{Evolution}
{{\partial P(C,t)} \over {\partial t}} = \sum_{C'} W(C' \to C) P(C',t)
- \sum_{C'} W(C \to C') P(C,t).
\end{equation}
The first sum represents the rate of flow, in
configuration space, of probability
into $C$ while the second sum corresponds to the outgoing
flow from this configuration. In a steady state the two terms
are equal, yielding zero net flow from any configuration.

Systems in thermal equilibrium\index{thermal
equilibrium} are characterised by an energy
function, or a Hamiltonian, $E(C)$. The steady state
distribution $P(C)$ is proportional to $e^{-E(C)/{k_B T}}$, where
$T$ is the temperature and $k_B$ is the Boltzmann constant.
Given an energy function $E(C)$ one can always find transition
rates $W(C \to C')$, such as the Metropolis rates,
which obey detailed balance. Here the two sums cancel term by term
\begin{equation}
\label{DetailedBalance}
W(C' \to C) P(C') = W(C \to C') P(C),
\end{equation}
for any pair of configurations $C$ and $C'$.

On the other hand dynamical systems are not defined by
an energy function but rather by transition rates.
When a system is not in thermal equilibrium, the resulting steady
is such that detailed balance
(\ref{DetailedBalance}) is not satisfied. We will basically
use this lack of detailed balance as a definition of nonequilibrium.

Given the dynamics of a system, namely the transition rates,
it is of interest to know whether or not detailed balance is satisfied.
Since, in general, the steady state distribution cannot be calculated,
a direct check of the detailed balance condition (\ref{DetailedBalance})
is not possible. Thus a criterion for existence of detailed balance which
is based directly on the transition rates and does not require
the knowledge of the steady state is highly desirable.
Such a criterion is provided by the following equations. Let
$C_1,C_2, \ldots ,C_k$ be a set of $k$ microscopic configurations.
A necessary and sufficient condition for the existence of detailed balance is
that for any such set one has
\begin{equation}
\label{Criterion}
W(1 \to 2) W(2 \to 3) \ldots W(k \to 1) = W(1 \to k) W(k \to k-1) \ldots W(2 \to 1),
\end{equation}
where for simplicity we have denoted $C_i$ by $i$.
It is easy to check that (\ref{Criterion}) is a necessary condition.
When detailed balance is satisfied one may replace
$W(i \to i+1) / W(i+1 \to i)$ by $P(i+1)/P(i)$, where P(i) is the 
steady state distribution with respect to which detailed balance
is satisfied. Using these relations (\ref{Criterion}) is easily verified.

To demonstrate that this is a sufficient condition as well,
we use (\ref{Criterion}) to derive the steady state distribution.
We start with an arbitrary configuration $1$ and denote its steady state
weight by $P(1)$. 
The weight of states $2$ which are directly connected with
$1$ (namely, for which $W(1 \to 2)>0$), may thus be defined using the
detailed balance relation, $P(2)=P(1) W(1 \to 2) / W(2 \to 1)$.
This process may then be repeated to define the weights of states directly
connected with states $2$ etc, until all microscopic configurations have
been reached. The weight of a 
microscopic configuration $k$ which may be reached
from $1$ via intermediate states $2,3, \ldots ,k-1$ is thus given by
\begin{equation}
\label{Pk}
P(k)= P(1) {{W(1 \to 2) \ldots W(k-1 \to k)} \over {W(k \to k-1) \ldots
W(2 \to 1)}}.
\end{equation}
For this procedure to be self-consistent one has to verify that any
path between configurations $1$ and $k$ yields the same $P(k)$.
It is a straightforward matter to show that this follows directly
from (\ref{Criterion}).

Thus, to demonstrate that a dynamical system defined by its transition rates
is not in thermal equilibrium it is sufficient to find a single path
in configuration space for which (\ref{Criterion}) is not satisfied.
This is usually quite easy  to check, making it a very useful criterion.
When (\ref{Criterion}) is not satisfied the system exhibits non-vanishing
probability currents between configurations 
which is an indication of the system not being in thermal 
equilibrium.

A simple prototypical model of driven systems, termed the `standard
model'\label{standard model}, was introduced by Katz {\it et al}
\cite{Katz,Zia}.  This is a driven lattice gas model defined on a
hypercubic lattice with periodic boundary conditions.  Each site $i$
is either occupied by a particle or is vacant, with $\sigma_i =0,1$
being the occupation number. In the absence of drive, an Ising
Hamiltonian is assumed
\begin{equation}
\label{Ising}
H = -J \sum_{\langle ij \rangle} \sigma_i \sigma_j,
\end{equation}
where the sum is over nearest neighbour (nn) sites $\langle ij
\rangle$. The evolution of the system is defined by Kawasaki dynamics,
allowing for particles to hop between nearest neighbour sites. Let $C$
and $C'$ be two configurations obtained from each other by an
interchange of a single pair of nn occupation numbers $\sigma_i$ and
$\sigma_j$. The transition rate between $C$ and $C'$ may be taken as
the Metropolis rate $W(C \to C') = w(\beta \Delta H)$, where $\Delta
H=H(C')-H(C)$ and $w(x)=\min (1,e^{-x})$. This dynamics leads to the
expected Boltzmann equilibrium distribution. In $d>1$ dimensions the
system exhibits the usual Ising transition from a homogeneous phase at
high temperatures to a phase separated state at low temperatures.

Introducing a driving field $E$ along one of the axes, the transition
rates are modified by adding a term $uE$ to $\Delta H$ where
\begin{equation}
\label{u}
u=-1 \; , \; 0 \; ,\; +1,
\end{equation}
for a hop along, transverse or opposite to the field direction,
respectively. The transition rates are thus given by
\begin{equation}
\label{DriveRates}
W(C \to C')= w(\beta ( \Delta H +uE)).
\end{equation}
Due to the periodic boundary conditions in the direction of the 
driving field, these rates do not obey detailed balance, and the
steady exhibits non-vanishing currents.

In spite of the simplicity of the model, no exact results for the
steady state properties are available (exact in one dimension).
Extensive numerical studies of this model in two dimensions
demonstrate that the phase separation transition which exists in
zero drive, persists for non-zero driving fields. At low temperatures
the system exhibits stripes of high density and low density
regions which are oriented along the field direction. These stripes coarsen
with time leading to a phase separated state.

\section{A\lowercase{symmetric} \lowercase{exclusion process 
in one dimension}}
\label{AEP}

In this section we consider the phase diagram of the one dimensional
`standard model' defined in the previous section for $J=0$.
Here the only interaction between the particles is
the hard-core interaction which prevents more than one particle from occupying the
same site. This process is called asymmetric simple exclusion process
\index{asymmetric simple  exclusion process} (ASEP).
Furthermore, we consider the limit $E \to \infty$,
called totally asymmetric simple exclusion process (TASEP). In this limit
particles are restricted to move only to the right, with no backward moves.
This model turns out to be sufficiently simple to allow for exact 
calculation of some of its steady state properties. In spite of
its simplicity, the model with open boundary conditions exhibits
a rather rich and complicated phase diagram, displaying both continuous and
discontinuous phase transitions (see below). This is clearly a direct consequence of
the fact that the dynamics is a nonequilibrium one. The model and many
variants of it have been a subject of extensive studies in recent years
\cite{MacDonald1,MacDonald2,Krug,DDM,DEHP,SD,Schutz,Haye,EvansSpeer,dGN}.
  
The dynamics of the model is defined as follows: at any given time
a pair of nn sites is chosen at random. If the occupation numbers of these sites
are $(+ \;0)$ an exchange is carried out
\begin{equation}
\label{Exchange}
+ \; 0 \to 0 \; + ,
\end{equation}
with rate $1$. All other configurations, remain unchanged.  Here and
in the following we interchangably use $1$ or $+$ to denote an
occupied site.  For periodic boundary conditions the system reaches a
trivial steady state in which all microscopic configurations have the
same weight. This may be verified by direct inspection of the master
equation (\ref{Evolution}).  It is easy to see that the number of
configurations to which a given configuration $C$ may flow is equal to
the number of configurations flowing into $C$.  To verify that this is
the case note that a microscopic configuration $C$ is composed of
alternating segments of $+'s$ and $0's$.  According to the dynamics
(\ref{Exchange}) $C$ may be exited when the rightmost particle in one
of the $+$ segments moves one step to the right. Thus the number of
configurations $C$ may flow into is equal to $\it l$, the number of
$+$ segments in this configuration. Similarly $C$ may be reached when
the leftmost particle in one of the $+$ segments hops into its
position. The number of configurationd flowing into $C$ is therefore
also equal to $\it l$.  Since all non-vanishing transition rates are
$1$, the state where all configurations have equal weights is
stationary. Therefore in the steady state the system exhibits no
correlations, apart from the trivial correlations arising from the
fact that the overall density of particles is fixed.

The steady state current $J$ is given by
\begin{equation}
\label{Current}
J= \langle \sigma_i (1- \sigma_{i+1}) \rangle ,
\end{equation}
where the brackets denote a statistical average with respect to the
steady state weights of the microscopic configuration. Since in the
steady state the system exhibits no correlations the current may be
written, in the large system limit, as
\begin{equation}
\label{Current1}
J= p (1- p),
\end{equation}
where $p_i = \langle \sigma_i \rangle $ is the density at site $i$,
and the index $i$ is omitted in (\ref{Current1}) since the average
density is homogeneous, independent of $i$. Equation (\ref{Current1}),
relating the current to the density is known as the fundamental
relation\index{fundamental relation} (or fundamental
diagram\index{fundamental diagram}).  The interesting feature in this
relation is that the current is not a monotonic function of the
density but rather it exhibits a maximum at $p = 0.5$. This feature is
a result of the hard-core interaction between the particles and it
affects rather drastically the steady state properties of the system
when open boundary conditions are considered.

\begin{figure}
\epsfxsize 10 cm
\hspace{3cm}\epsfbox{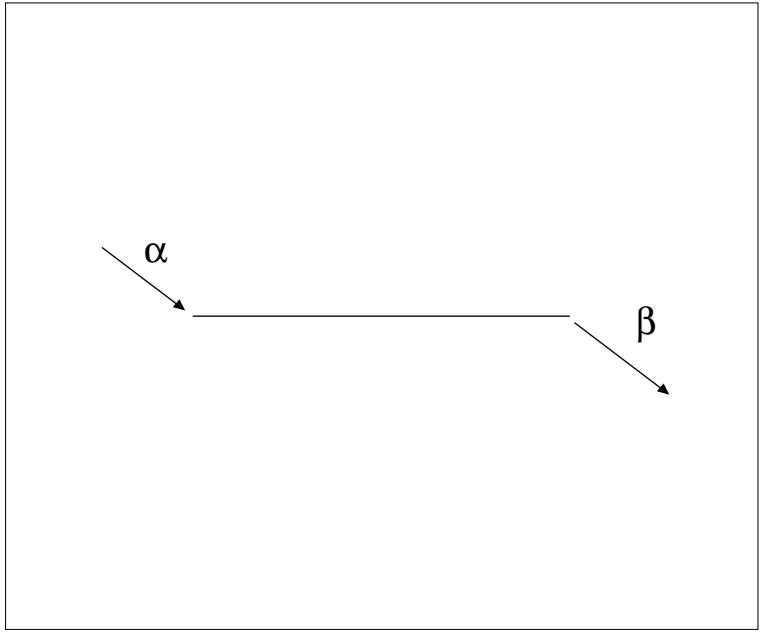}
\vspace{0.5cm}
\caption{A schematic representation of the totally asymmetric exclusion
process with open boundary conditions.}
\label{openbc}
\end{figure}

We now turn to the model with {\it open} boundary conditions.
Here, particles are introduced
into the system at the left end, they move through the bulk according to the
conserving dynamics (\ref{Exchange}), and leave the system at the right end.
To be more specific, at the left boundary $(i=1)$ the move
\begin{equation}
\label{Right}
0 \to +  ,
\end{equation}
is carried out with a rate $\alpha$. Similarly, at the right boundary $(i=N)$
one takes 
\begin{equation}
\label{Left}
+ \to 0  ,
\end{equation}
with a rate $\beta$. For a schematic representation of the model 
see Figure~\ref{openbc}.
The overall dynamics is non-conserving. Particles
are conserved in the bulk but are not conserved at the boundaries.
Unlike the case with periodic boundary conditions, the steady state
distribution is not trivial, and correlations between the densities
at different sites do not vanish. However, far from the boundaries
the distribution function is expected to be well approximated by the
homogeneous one, suggesting that correlations are small. 

We are interested in the steady state of this model for given rates
$\alpha$ and $\beta$. For large $\alpha$ and small $\beta$, namely for a
large feeding rate and a small exit rate the overall density is 
expected to be high. On the other hand for small $\alpha$ and large
$\beta$ the density is expected to be low. In addition, for a large
system, where away from the boundaries the local density is expected
to vary very slowly, the fundamental relation (\ref{Current1}) is
expected to hold locally. Thus the current in the system cannot exceed a 
maximal current, as suggested by (\ref{Current1}). These features
yield a rather rich phase diagram, as $\alpha$ and $\beta$ are varied.

We start by considering the phase diagram in the mean field approximation
\cite{MacDonald1,MacDonald2,DDM}.
Since correlations in this system are expected to be vanishingly small
away from the boundaries, this approximation is expected to yield
a rather accurate phase diagram. In fact it turns out that the
phase diagram obtained in this way is exact.

To derive the mean field
equations we note that the current $J_{i,i+1}$ between sites $i$ and
$i+1$ is given by $\langle \sigma_i (1-\sigma_{i+1}) \rangle $
for $i=1, \ldots ,N-1$. In addition the currents at the two
ends are given by $J_0=\alpha  \langle 1-\sigma_1 \rangle $ and
$J_N=\beta \langle \sigma_N \rangle $.
Neglecting correlations one finds that in the steady state,
where all currents are equal, the following equations
have to be satisfied:
\begin{equation}
\label{MFcurrent}
J=\alpha(1- p_1) = p_1 (1-p_2)= \ldots =p_{N-1}(1-p_N) = \beta p_N .
\end{equation}
Solving these equations for $J,p_1, \ldots ,p_N$ the density profile in
the steady state and the current are obtained.

It is instructive to consider these equations in the continuum limit.
Replacing  $p_i$ by $p(x)$ in (\ref{MFcurrent}) with $0 \le x \le L$
yields the bulk current 
\begin{equation}
\label{ContinuumJ}
J(x)=p(1-p) -D {{\partial p} \over {\partial x}} ,
\end{equation}
where $D$ is the diffusion constant which, by rescaling $x$, may be
taken as $1$. In this expression the first term represents the drive
while the second term is the ordinary diffusion current. The evolution
of the system is governed by the continuity equation
${\partial p }/ {\partial t} = - {\partial J}/ {\partial x}$, together
with the boundary conditions $J(0)=\alpha (1-p(0))$ and $J(L)=\beta p(L)$.
In the steady state the current $J$ in (\ref{ContinuumJ})
is independent of $x$ yielding a density profile which has one of
the two following forms 
\begin{eqnarray}
p(x) &=& 0.5 + v \tanh[v(x-x_0)] \nonumber \\
p(x) &=& 0.5 +v \coth[v(x-x_0)] ,
\end{eqnarray}
where $v^2=1/4-J$. The two parameters $x_0$ and $v$ (or alternatively
the current $J$) are determined by the two boundary conditions, and are
thus related to $\alpha$ and $\beta$.

\begin{figure}
\epsfxsize 10 cm
\hspace{3cm}\epsfbox{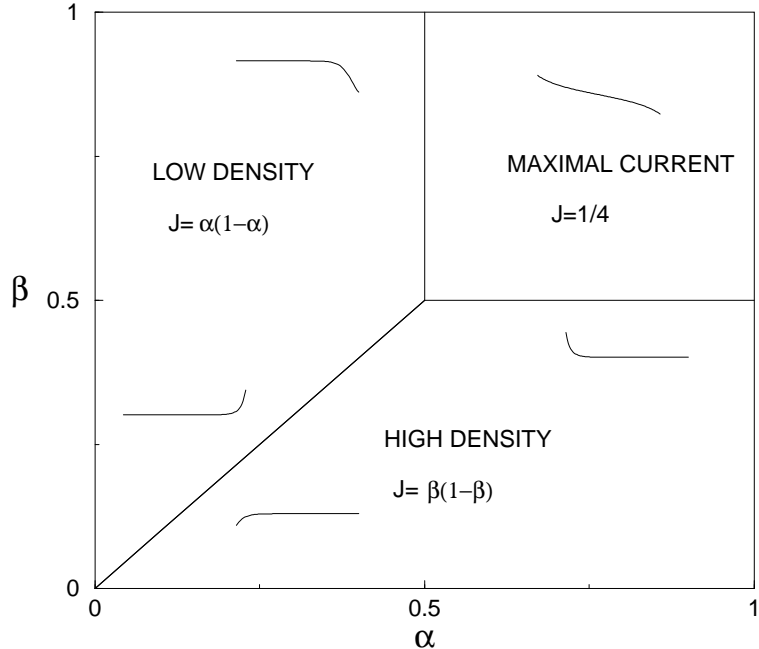}
\vspace{0.5cm}
\caption{The $(\alpha , \beta)$ phase diagram of the totally
asymmetric exclusion process exhibiting low density, high density and
maximal current phases. The thick line represents a first order
transition while thin lines correspond to continuous transitions.
Schematic density profiles in the various phases are given.}
\label{phasediagram}
\end{figure}

By matching the boundary conditions the density profiles and the 
current are obtained. The resulting phase diagram is
given in Figure~\ref{phasediagram}. The system is found to exhibit three
distinct phases in the limit of large length $L$:

\begin{itemize}
\item Low density phase for which the bulk density is smaller than $0.5$
with $x_0 = O(L)$.
The density profile is basically flat, except for a small region
near the right end. In this phase $p(0)=\alpha$ and $J=\alpha (1-\alpha)$.
It exists for $\alpha < \beta$ and $\alpha < 1/2$.

\item High density phase with bulk density larger than $0.5$ with $x_0=-O(L)$.
The density profile
is flat except at a small region near the left end. Here $p(L)=1-\beta$
and $J=\beta (1-\beta)$. This phase exist for $\beta < \alpha$
and $\beta < 1/2$.
The two phases co-exist on the line $\alpha = \beta < 1/2$.

\item A maximal current phase in the region
$\alpha > 1/2$ and $\beta > 1/2$. Here the bulk density is 1/2
exhibiting structures at both ends of the system. These structures decay
algebraically as $1/x$ when moving away from the ends. In this
phase the current is maximal, namely $J=1/4$.
\end{itemize}

The phase diagram exhibits
a first order line on which the high density and the low density
phases coexist and two second order lines separating these
phases from the maximal current phase. Typical schematic density
profiles in the various phases are also given in Figure~\ref{phasediagram}.

In the mean field approximation fluctuations are neglected, and therefore
by itself this analysis may not serve as a demonstration that phase transitions
do take place in $1d$ away from thermal equilibrium. In fact
mean field approximation yields phase transitions in equilibrium $1d$
systems, where they are known not to exist. It is therefore 
important to examine the role of fluctuations in this driven system
and demonstrate that indeed the phase transitions found within the mean
field approximation remain when fluctuations are taken into account.
This has indeed been demonstrated for the TASEP \cite{DDM,DEHP,SD}.
A method which goes beyond the mean field approximation and allows
exact calculations of steady state properties of some driven
$1d$ models is described in the next section.

\section{M\lowercase{atrix} \lowercase{method}}
\label{Matrix}

A matrix method\index{matrix method} for calculating some steady state properties
of the TASEP was introduced a few years ago \cite{DEHP}.
The method has
since then been generalised and applied to other models of driven
systems. In the following we briefly outline the method as applied to
the TASEP with open boundary conditions described above.

We are interested in calculating the steady state distribution function
$P(\sigma_1, \ldots ,\sigma_N)$. In the matrix method one tries to express the
distribution function by a matrix element of a product of particular matrices.
Let $D$ and $E$ be two square matrices and $ \langle W|$ and $|V \rangle $ be two vectors.
For any given configuration, $(\sigma_1, \ldots ,\sigma_N)$ one 
considers the matrix product in which each occupation number $\sigma_i$ 
is replaced by either a matrix $D$ or a matrix $E$ depending on whether
it is $1$ or $0$, respectively. The key question is whether
one can find matrices $D$ and $E$ and vectors $ \langle W|$ and $|V \rangle $
such that $P(\sigma_1, \ldots ,\sigma_N)$ is proportional to
the $ \langle W|,|V \rangle $ matrix element of this product. Within this
representation the distribution function may be written as
\begin{equation}
\label{MatrixProduct}
P(\sigma_1, \ldots ,\sigma_N) \propto  \langle W| \prod_{i=1}^N [\sigma_i D
+ (1-\sigma_i) E]|V \rangle . 
\end{equation}
{\it A priori}, it is not at all clear that such representation is available.
However, if such representation exists, it may yield
a straightforward (though sometimes tedious) way for calculating the
distribution function. For example, the density at, say, site $i=1$
may be expressed as
\begin{equation}
\label{eq:p1}
p_1={1\over Z_N} \sum_{\lbrace X_i \rbrace}  \langle W|D X_2 \ldots X_N|V \rangle , 
\end{equation}
where $X_i = D, E$, and the normalization factor $Z_N$ is given by
\begin{equation}
Z_N= \sum_{\lbrace X_i \rbrace}  \langle W|X_1 X_2 \ldots X_N|V \rangle  = 
\langle W|C^N|V \rangle , 
\end{equation}
with
\begin{equation}
C=D+E \;\;\; .
\end{equation}
Thus, we may rewrite (\ref{eq:p1}) as
\begin{equation}
p_1 = {1\over Z_N} \langle W|D C^{N-1}|V \rangle \; .
\end{equation}
Densities at other sites and density-density correlation 
functions may similarly be expressed by other matrix elements.

The main question at this point is how to find matrices and 
vectors such that (\ref{MatrixProduct}) holds. To this end we
consider the local currents in the system. Using the matrix representation, the
current between sites $i$ and $i+1$ $(i=1, \ldots ,N-1)$ may be expressed as
\begin{equation}
J_{i,i+1}={1 \over Z_N} \langle W|C^{i-1} D E C^{N-i-1}|V \rangle , 
\end{equation}
and the currents at the two ends 
\begin{eqnarray}
J_0 &=& {\alpha \over Z_N}  \langle W|E C^{N-1} |V \rangle \nonumber \\
J_N &=& {\beta \over Z_N}  \langle W|C^{N-1} D|V \rangle . 
\end{eqnarray}
In the steady state all currents are equal. Taking matrices $D,E$ and vectors
$ \langle W| , |V \rangle $ which satisfy
\begin{eqnarray}
\label{MatrixConditions}
&&DE=D+E \; \; (=C) \nonumber \\
&&\alpha \langle W\vert E = \langle W | \\
&&\beta D \vert V \rangle  = \vert V \rangle  , \nonumber
\end{eqnarray}
guarantee that all currents are equal, with
\begin{equation}
J={Z_{N-1} \over Z_N} .
\end{equation}
The question is whether these relations (\ref{MatrixConditions}) are sufficient to guarantee that the
resulting distribution (\ref{MatrixProduct}) is a steady state. For
(\ref{MatrixProduct}) to be a steady state one has to make sure that
\begin{equation}
\label{Stationarity}
{\partial P(\sigma_1, \ldots ,\sigma_N)} / {\partial t} =0 ,
\end{equation}
for each of the $2^N$ microscopic configurations. The relations
(\ref{MatrixConditions}) only directly guarantee that $N+1$ of these $2^N$
equations are satisfied. This may suggest that (\ref{MatrixConditions})
may not be sufficient to guarantee that the steady state distribution is
given by (\ref{MatrixProduct}). However it can be shown, by direct inspection
of Equations (\ref{Stationarity}) that (\ref{MatrixConditions}) yields the
steady state of the system. 

The problem is thus reduced to first finding matrices and vectors which
satisfy (\ref{MatrixProduct}), and then calculating some matrix elements
to obtain, for example, the current $J$. It is straightforward to show
that for $\alpha + \beta \ne 1$ the matrices which satisfy
(\ref{MatrixProduct}) have to be of infinite order. Such matrices
have been found and the current and density profiles and other
correlation functions have been calculated \cite{DEHP}.
The resulting phase diagram
coincides with that obtained by the mean field approximation, although 
the density profiles are different. For example, in the algebraic, maximal
current, phase the local density decays to the bulk density
like $1 / \sqrt{x}$ at large distances from
the boundary, unlike the mean field result which yields a $1 / x$
profile.

The matrix method proved to be very powerful in yielding steady state properties
of TASEP dynamics. It has been applied and generalised to study partially asymmetric
exclusion processes (ASEP) \cite{Sandow,Sasamoto,EvansPartial} and models
with more than one type of particles
\cite{SecondClass,EvansSecond,ArndtAlgebra,Bridge1}. In addition, replacing
matrices by tensors proved to be useful in some cases \cite{DEM,ABC}.
However the method is restricted to one dimension. It
is not standard in the sense that it can not be applied
to any dynamical model. Moreover, there is no simple way to tell a priori
whether or not it may be applicable for a specific model. 

\section{S\lowercase{pontaneous} \lowercase{symmetry breaking
 in one dimension}}
 \label{SSB}

In this section we consider a simple dynamical model which exhibits
{\it spontaneous symmetry breaking}\index{spontaneous symmetry
breaking} (SSB) in $1d$.

The model may be pictorially described in the following way:
consider a narrow bridge connecting two roads. Cars travelling
on the bridge in opposite directions do not block each other,
although they may slow the traffic flow in both directions.
We assume that the two roads leading to the bridge
from both sides
are statistically identical. Namely the arrival rates of cars
at the two ends of the bridge are the same. This system clearly
has a right-left symmetry. Thus if this symmetry is not spontaneously
broken one would expect that the long time average of the current
of cars travelling to the right would be the same as that of
cars travelling to the left. The question is whether the bridge is
capable of exhibiting breaking of the right-left symmetry, and
spontaneously turning itself into a `one-way' street,
where the current in one direction is larger than the current in the
other direction. It turns out that this may indeed take place 
in the limit of a long bridge, demonstrating that SSB may 
take place in $1d$ nonequilibrium systems.

To model the `bridge' problem\index{bridge problem} we generalise the TASEP discussed in the
previous section \cite{Bridge1,Bridge2,RittenbergBridge}.
We consider a $1d$ lattice
of length $N$. Each lattice point may be occupied by either a $(+)$
particle (positive charge) moving to the right, a $(-)$ particle
(negative charge) moving to the left or by
a vacancy $(0)$. In addition positive (negative) charges are supplied at
the left (right) end and are removed at the right (left) end of the
system.

\begin{figure}
\epsfxsize 10 cm
\hspace{3cm}\epsfbox{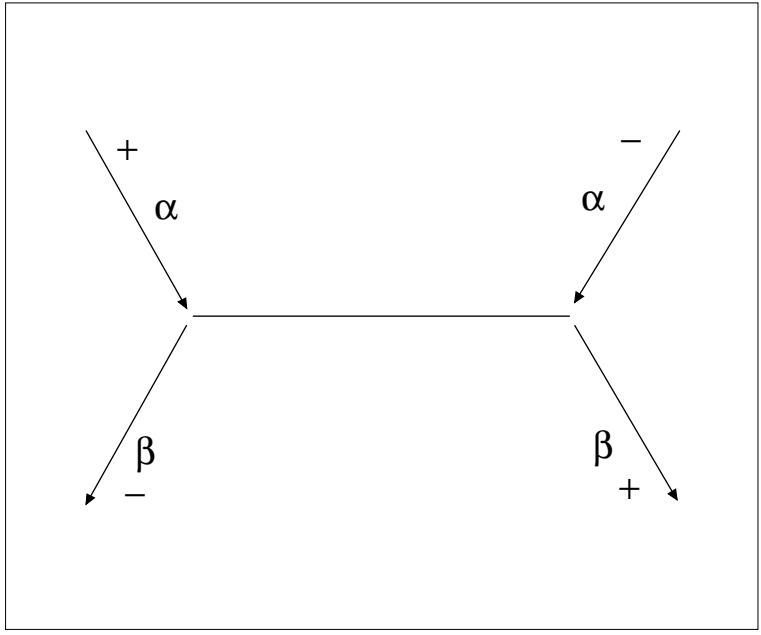}
\caption{A schematic representation of the input and output rates
of the "bridge' model.}
\label{bridge}
\end{figure}

The dynamics of the model is defined as follows: at each time
step a pair of nearest neighbour sites is chosen and an exchange
process is carried out
\begin{equation}
+ \; 0 \to 0 \; + , \;\;\;\; 0 \; - \to - \; 0 , \;\;\;\; + \; - \to - \; +  ,
\end{equation}
with rates $1$, $1$ and $q$, respectively. Furthermore, at the two
ends particles may be introduced or removed. At the left boundary
$(i=1)$ the processes
\begin{equation}
0 \to + \;\;\;\; , \;\;\;\; - \to 0 ,
\end{equation}
take place with rates $\alpha $ and $\beta $, respectively. Similarly, at the
right boundary $(i=N)$, one has
the processes
\begin{equation}
0 \to - \;\;\;\; , \;\;\;\; + \to 0 ,
\end{equation}
with rates $\alpha$ and $\beta$, respectively (see Figure~\ref{bridge}).
In the `bridge' language the boundary terms may be viewed as 
traffic lights which control the feeding and exit rates at the two ends.

Since the parameters $\alpha$ and $\beta$ are the same on both ends
of the systems
the dynamics obviously possesses a right-left symmetry. The question of
interest is whether or not this symmetry is preserved in the steady state.
Clearly, for small $\alpha$ and large $\beta$ the density of particles
in the system is expected to be low, the two types of particles do not
block each other, and the steady state is expected to be symmetric.
On the other hand for $\beta$ much smaller than $\alpha$, particles
are blocked in the system, the density is high and it is possible that
symmetry breaking takes place. 

We start by considering the mean field approximation.
It is straightforward to derive the mean field equations for the steady state.
They take the form
\begin{eqnarray}
J_+ &=& p_i [1-p_{i+1} - (1-q)m_{i+1}] \nonumber \\
J_- &=& m_{i+1} [1-m_i - (1-q)p_i]  , 
\end{eqnarray} 
for $i=1, \ldots ,N-1$, where $p_i$ and $m_i$ are the densities of the $(+)$ and
$(-)$ particles at site $i$, respectively, and $J_+$ and $J_-$ are the currents
of the positive and negative particles, respectively. In addition to the
bulk equations one has four other equations for the currents at the 
boundaries
\begin{eqnarray}
J_+ &=& \alpha (1-p_1 - m_1)=\beta p_N \nonumber \\
J_- &=& \beta m_1 = \alpha (1-p_N - m_N) .
\end{eqnarray}
These $(2N+2)$ equations may be solved numerically for
$(p_1, \ldots ,p_N;m_1, \ldots ,m_N;J_+,J_-)$ to yield the $(\alpha, \beta)$
phase diagram of the model. It is found that for large $\beta$
the steady state is symmetric (with $J_+ = J_-$) while for
small $\beta$ the two currents are unequal in the steady state.

The matrix method discussed in the previous section has been 
generalised and applied to this model \cite{Bridge1}. However it turned out
that a self-consistent matrix representation could be
found for this model only for $\beta =1$ or in the limit
$\alpha \to \infty$. The limit $\alpha \to \infty$ is trivially
mapped on the single species TASEP model. For $\beta =1$
a phase transition is found although no spontaneous symmetry
breaking takes place. According to mean field SSB is expected only
at much lower exit rates $\beta$.

\begin{figure}
\epsfxsize 10 cm
\hspace{3cm}\epsfbox{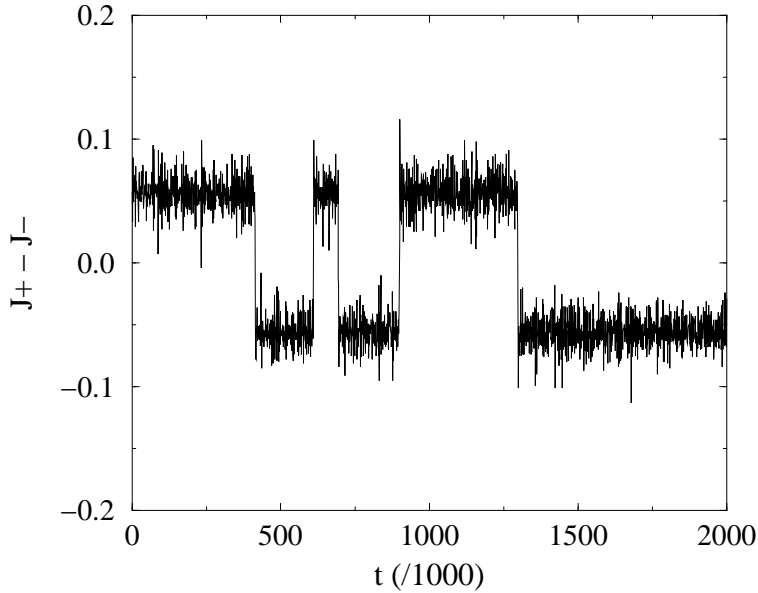}
\caption{The time evolution of the current difference in the broken
symmetry phase for ($\alpha =q =1, \beta =0.15, N=80$). Time is measured
in units of Monte Carlo sweeps. Flips between the two symmetry related
states are clealy seen.}
\label{current}
\end{figure}

To demonstrate that the non-symmetric state found in the mean field
approximation for small $\beta$ survives fluctuations,
numerical simulations of the dynamics have been carried out and the 
current difference $J_+ - J_-$ has been measured as a function of time
for small $\beta$, where the mean field approximation predicts a 
broken symmetry phase \cite{Bridge1}.
A typical time evolution of the current
difference for a system of size $N=80$ is given in Figure~\ref{current}.
%
The figure
suggests that the system flips between two macroscopic states: one 
with a positive net current and the other with negative net current.
In the first case the system is predominantly loaded with positive
charges moving to the right while in the second case it is loaded
with negative charges moving to the left. This time course is
characterised by a time-scale $\tau (N)$ which measures the average
time between flips. Clearly, when averaged over time,
the current difference vanishes, yielding a symmetric state.
This is to be expected since we are dealing with a finite system,
and one certainly does not expect SSB to take place in a finite
system. The question is how does the system behave in the thermodynamic
limit $N \to \infty$, and particularly how does $\tau (N)$ grow for
large $N$. Numerical simulations suggest that $\tau$ 
grows exponentially with $N$. This means that the probability 
of a flip is negligibly small in a large system, and thus SSB takes
place.

\begin{figure}
\epsfxsize 10 cm
\hspace{3cm}\epsfbox{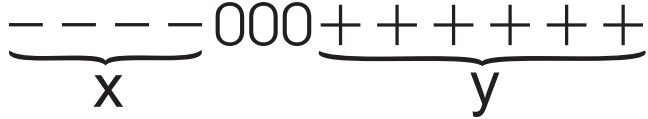}
\caption{A typical microscopic configuration of the "bridge" in the
broken symmetry state for small $\beta$.
Usually $x$ is a small number of $O(1)$ and $y$
is of $O(N)$ or vise versa.}
\label{xytoy}
\end{figure}

In order to gain some insight into the flipping process we consider
the limit of very small $\beta$ \cite{Bridge2,ArndtToy}.
In this limit, particles leave the system
at a very small rate, and the system is filled with either positive charges
moving to the right or negative charges moving to the left. Starting with
a positively charged system, one would like to understand the mechanism by which
a system of finite length flips into a negatively charged one.
The evolution in the small $\beta$ limit may be described as follows:
with rate $\beta$ a positive charge leaves the system at the right end.
The vacancy created at this end may either move to the left with velocity $1$
or may be filled with a negatively charged particle which in turn moves to
the left with velocity $q$. When the negative charge reaches
the other end of the system it is
delayed for a while, but eventually leaves the system
in time of order $ 1 / \beta$.
During this time, other negative charges may arrive at
the left end forming a small blockage of negative charges. A typical
configuration is given in Figure~\ref{xytoy}.
%
 It is composed of a segment of $x$ negative
charges at the left, another segment of $y$ positive charges at the
right and in between a segment of $N{-}x{-}y$ vacancies. This configuration is
denoted by $(x,y)$. In the small $\beta$ limit other configurations,
for example those in which vacancies are present inside the charged segments,
do not play a role in the global dynamics and may be neglected.

\begin{figure}
\epsfxsize 9 cm
\hspace{4cm}\epsfbox{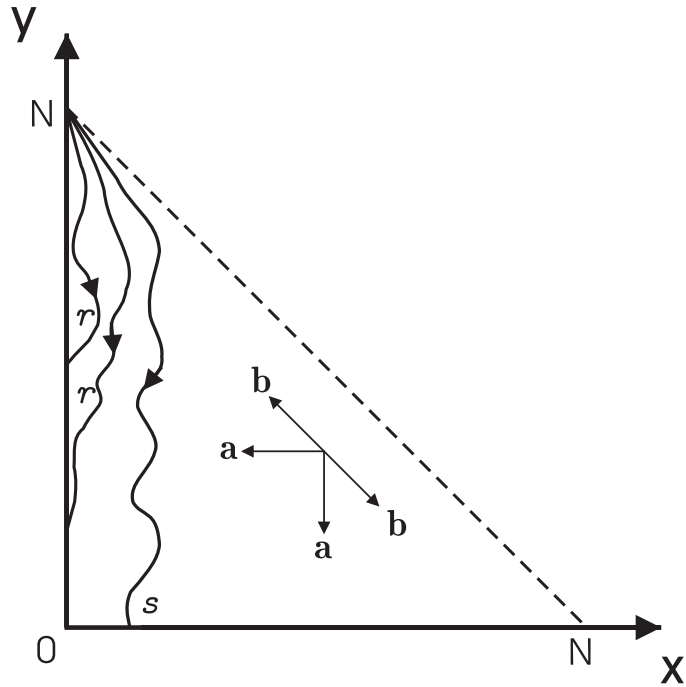}
\vspace{0.5cm}
\caption{A representation of the dynamics of the "bridge" in the limit
of small $\beta$ as a biased diffusion process on a square lattice.
The elementary moves and their rates are indicated.
Starting with a positively charged system, namely at $(N,0)$, possible
trajectories are given. Typical walks, $r$ , end on the $x=0$ axis.
Walks, $s$ , which end on the $y=0$ axis are rare, and they correspond 
to a flip. }
\label{bias}
\end{figure}

The dynamics restricted to $(x,y)$ configurations is rather simple.
It may be viewed as the dynamics of a single particle diffusing on
a square lattice performing the following elementary moves:
\begin{eqnarray}
\label{Toy}
(x,y) & \stackrel{\textstyle b}{ \to }& (x+1,y-1)  \nonumber \\
(x,y) &\stackrel{\textstyle  a}{ \to }& (x,y-1)    \nonumber \\
(x,y) &\stackrel{\textstyle  b} { \to }& (x-1,y+1)    \\
(x,y) &\stackrel{\textstyle  a}{ \to }& (x-1,y)  ,   \nonumber
\end{eqnarray}
where
\begin{equation}
a={1 \over {2(1+\alpha)}} \;, \;\;\;\;
b={{\alpha}\over {2 (1+\alpha)}} ,
\end{equation}
are the rates of the various moves.
Here the first two moves correspond to a positive charge
leaving the system at the right end and being replaced by a negative charge
(vacancy), respectively. Similarly the last two moves correspond to a
negative charge leaving the system at the left end and replaced by a
positive charge (vacancy), respectively. A schematic representation of
this process is given in Figure~\ref{bias}.

The biased diffusion process takes
place as long as the particle stays within the triangle $(x \ge 0 \;,\; y \ge 0 \;,\;
x+y \le N)$. When it reaches the boundary of the triangle, for example
$x=0$ the negative charge blockage at the left end disappears and on
a very short time scale (as compared with $1/\beta$) the system is filled
with positive charges from the left end, moving to $(0,N)$. The evolution
of a positively charged system is thus represented by a random walk starting
at $(0,N)$ with elementary steps defined by (\ref{Toy}). Due to the bias
of these elementary steps, a typical walk for large $N$ ends on the $x=0$
axis. Once it reaches this axis it moves back to $(0,N)$ and the process
starts again. This 
process repeats itself until the diffusing particle performs a walk
which starts at $(0,N)$ and ends on the $y=0$ axis without touching
the $x=0$ axis while diffusing. When this happens the blockage of positive
charges at the right end is removed, the system is rapidly filled with
negative charges moving to the other end of the triangle $(N,0)$.
This corresponds to a flip. The probability of such a walk taking
place has been calculated, yielding the following flipping time
\index{flipping time} \cite{Bridge2}:
\begin{equation}
\tau (N) = {C \over \beta} N^{3/2} e^{\kappa N}  ,
\end{equation}
where $C$ is a constant and
\begin{equation}
\kappa = \ln \Big[{2 \over a} \Big(1-a-\sqrt{(1-a)(1-2a)}\; \Big)\Big]  .
\end{equation}
The exponential flipping time is a direct result of the  fact that
the random walk corresponding to the evolution of the model is
biased. It takes an exponentially long time to reach a distance of
order $N$ against a bias.

A very interesting question is related to the behaviour of
nonequilibrium systems with spontaneous symmetry breaking when an
{\it external symmetry breaking field}\index{external symmetry
breaking field} is introduced. In thermal equilibrium a symmetry
breaking field makes the phase unfavoured by the field metastable or
even unstable (when the field is large).  For example, a positive
magnetic field applied to an ordered ferromagnetic Ising system,
removes the degeneracy between the two magnetic states.  Only the
state with positive net magnetization remains stable.  The two
magnetic states coexist only at zero field. This is a direct
consequence of the Gibbs phase rule.

It is known that in nonequilibrium systems, this is not necessarily
the case \cite{Toom,Geoff}.
Namely when a symmetry breaking field is applied, the
state unfavoured by the field may stay as a stable thermodynamic
state. This is in violation of the Gibbs phase rule,
which does not hold in nonequilibrium.

\begin{figure}
\epsfxsize 6 cm
\hspace{5.5cm}\epsfbox{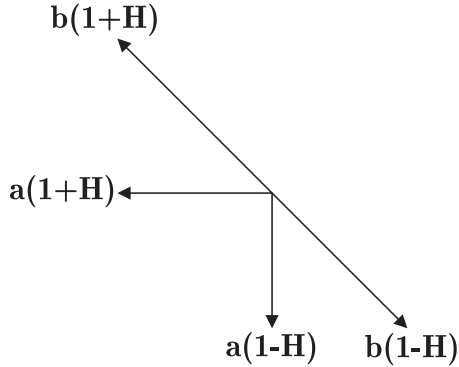}
\vspace{0.5cm}
\caption{The elementary moves of the biased diffusion process corresponding
to the small $\beta$ limit when a symmetry breaking field $H$ is present.}
\label{diffusion1}
\end{figure}

The `bridge' model described in this section provides a clear
example for this behaviour \cite{Bridge2}.
To demonstrate this point we introduce
a symmetry breaking field by imposing boundary conditions
which favour, say, the positively charged state, thus explicitly
breaking the symmetry. More specifically, we consider
an exclusion model where, instead of having boundary rates
$\alpha , \beta$ for both types of particles,
we take $\alpha , \beta_+$ for the positive charges and $\alpha , \beta_-$
for the negative charges. The two exit rates are taken to be of the form
$\beta_{\pm} = \beta (1 \mp H)$, where $0 < H < 1$ is the symmetry breaking
field, favouring the positively charged state. The analysis presented above
in the limit $\beta \to 0$ may be repeated for non-vanishing field
$H$ and the stability of the two phases may be analysed. Here again
the dynamics is reduced to a diffusion process of the type (\ref{Toy})
but with modifies rates (see Figure~\ref{diffusion1}).
%
 Clearly the positively charged
state is stable since it is favoured by the field.
The question is whether
the negatively charged state is stable when the field $H$ is 
non-vanishing. To examine this problem, we start with a negatively
charged system $(N,0)$ and consider a random walk defined by the diffusion
process. It is easy to see that as long as
$H < a /(1-a)=1/(1 + 2 \alpha)$
the walk is biased in the negative $y$ direction, yielding
a flipping time exponential in the system size. Thus the negatively
charged state is {\it stable} even when it is unfavoured by
the symmetry breaking field.

\section{P\lowercase{ase} \lowercase{separation} \lowercase{in}
 \lowercase{one} \lowercase{dimension}}
 \label{PhaseSeparation}

A phenomenon closely related to spontaneous symmetry breaking is that
of phase separation\index{phase separation}.  In $1d$ equilibrium
systems with short range interactions phase separation does not take
place, and therefore no liquid-gas like transition is expected. The
density of particles in such a system is thus macroscopically
homogeneous.

Recent studies have shown that driven systems may exhibit phase
separation in $1d$ even when the system is governed
by local dynamics \cite{ABC,Ramaswamy1,Ramaswamy2,RittenbergABC}.
Several models have been introduced to 
demonstrate this behaviour. In these models more than one type of 
particles is needed for phase separation to take place.
In the following we consider in some detail
one of these models, and analyse the mechanism leading to
nonequilibrium phase separation \cite{ABC}.

The model is defined on a $1d$ lattice of length $N$ with periodic
boundary conditions. Each site is occupied by either an $A$, $B$, or
$C$ particle. The evolution is governed by random sequential dynamics
defined as follows: at each time step two neighbouring sites are chosen
randomly and the particles of these sites are exchanged according to
the following rates
\begin{equation}
\label{eq:dynamics}
\begin{picture}(130,37)(0,2)
\unitlength=1.0pt
\put(36,6){$BC$}
\put(56,4) {$\longleftarrow$}
\put(62,-1) {\footnotesize $1$}
\put(56,8) {$\longrightarrow$}
\put(62,14) {\footnotesize $q$}
\put(80,6){$CB$}
\put(36,28){$AB$}
\put(56,26) {$\longleftarrow$}
\put(62,21) {\footnotesize $1$}
\put(56,30) {$\longrightarrow$}
\put(62,36) {\footnotesize $q$}
\put(80,28){$BA$}
\put(36,-16){$CA$}
\put(56,-18) {$\longleftarrow$}
\put(62,-23) {\footnotesize $1$}
\put(56,-14) {$\longrightarrow$}
\put(62,-8) {\footnotesize $q$}
\put(80,-16){$AC$.}
\end{picture}
\end{equation}
\vspace{0.1in}

\noindent The rates are cyclic in $A$, $B$ and $C$ and conserve the number of
particles of each type $N_A,N_B$ and $N_C$, respectively.\

For $q=1$ the particles undergo symmetric
diffusion and the system is disordered. This is expected since this
is an equilibrium steady state. However for $q \neq 1$ the
particle exchange rates are biased. We will show that in this case
the system evolves into a phase separated state in the 
thermodynamic limit.

To be specific we take $q<1$, although the analysis may trivially
be extended for any $q \ne 1$. In this case the bias drives, say, an
$A$ particle to move to the left inside a $B$ domain, and to the right
inside a $C$ domain. Therefore, starting with an arbitrary initial
configuration, the system reaches after a relatively short transient
time a state of the type $\ldots AABBCCAAAB \ldots$ in which $A,B$ and
$C$ domains are located to the right of $C$, $A$ and $B$ domains,
respectively.  Due to the bias $q$, the domain walls $\ldots AB
\ldots$, $\ldots BC \ldots$, and $\ldots CA \ldots$, are stable, and
configurations of this type are long-lived. In fact, the domains in
these configurations diffuse into each other and coarsen on a time
scale of the order of $q^{-l}$, where $l$ is a typical domain size in
the system.  This leads to the growth of the typical domain size as $(
\ln t)/\vert\ln q \vert$. Eventually the system phase separates into
three domains of the different species of the form $A \ldots AB \ldots
BC \ldots C$. A finite system does not stay in such a state
indefinitely.  For example, the $A$ domain breaks up into smaller
domains in a time of order $q^{-min \lbrace N_B,N_C \rbrace}$. In the
thermodynamic limit, however, when the density of each type of
particle is non vanishing, the time scale for the break up of
extensive domains diverges and we expect the system to phase
separate. Generically the system supports particle currents in the
steady state. This can be seen by considering, say, the $A$ domain in
the phase separated state. The rates at which an $A$ particle
traverses a $B$ ($C$) domain to the right (left) is of the order of
$q^{N_B}$ ($q^{N_C}$). The net current is then of the order of
$q^{N_B}-q^{N_C}$, vanishing exponentially with $N$. This simple
argument suggests that for the special case $N_A=N_B=N_C$ the current
is zero for any system size.

The special case of equal densities $N_A=N_B=N_C$
provide very interesting insight into the mechanism leading
to phase separation. We thus consider it in some detail.
Examining the dynamics for these densities, one finds that
it obeys {\it detailed balance} with respect to some distribution
function. Thus in this case the model is in fact in thermal 
equilibrium. It turns out however that although the dynamics
of the model is {\it local} the effective Hamiltonian
corresponding to the steady state distribution has {\it long
range interactions}, and may thus lead to phase separation.
This particular mechanism is specific for equal densities.
However the dynamical argument for phase separation given above
is more general, and is valid for unequal densities as well.
 
In order to specify the distribution function for equal densities,
we define a local occupation variable $\lbrace X_i \rbrace
= \lbrace A_i,B_i,C_i \rbrace$, where $A_i$, $B_i$ and $C_i$ are equal
to one if site $i$ is occupied by particle $A$, $B$ or $C$
respectively and zero otherwise. The probability of finding the 
system in a configuration $\lbrace X_i \rbrace$ is given by
\begin{equation}
\label{eq:weight}
W_N(\{X_i\}) = Z_N^{-1}q^{{\cal H}(\lbrace X_i \rbrace)}  .
\end{equation}
where $\cal H$ is the Hamiltonian 
\begin{equation}
\label{Hamiltonian}
{\cal H}(\{ X_i \})=
\sum_{i=1}^N \sum_{k=1}^{N-1}
(1 - \frac{k}{N}) (C_i B_{i+k} + A_i C_{i+k} + B_i A_{i+k}) - (N/3)^2  ,
\end{equation}
and the partition sum is given by $ Z_N=\sum q^{{\cal H} (\lbrace
X_i \rbrace)}$. The value of the site index $(i+k)$ in (\ref{Hamiltonian})
is taken modulo $N$. In this Hamiltonian the interaction between
particles is long range, growing linearly with the distance between
the particles.

In order to verify that the dynamics (\ref{eq:dynamics}) obeys detailed
balance with respect to the distribution function
(\ref{eq:weight},\ref{Hamiltonian}) it is useful to note that the energy
of a given configuration may be evaluated in an alternate way.
Consider the fully phase separated state
\begin{equation}
\label{Groundstate}
A\ldots AB\ldots BC\ldots C
\end{equation}
The energy of this configuration is
$E=0$, and, together with its translationally relates configurations,
they constitute the $N-$fold degenerate  ground state of the system.
We now note that nearest neighbour (nn) exchanges $AB \rightarrow BA,
BC \rightarrow CB$ and $CA \rightarrow AC$ cost one unit of energy each,
while the reverse exchanges result in an energy gain of one unit.
The energy of an arbitrary configuration may thus be evaluated
by starting with the ground state and performing nn
exchanges until the configuration is reached, keeping track of the
energy changes at each step of the way. This procedure for obtaining
the energy is self consistent only when the densities of the
three species are equal. To examine self consistency of this procedure
consider, for example, the ground state (\ref{Groundstate}), and move
the leftmost particle $A$ to the right by a series of nn exchanges until
it reaches the right end of the system. Due to translational invariance,
the resulting configuration should have the same energy as 
(\ref{Groundstate}), namely $E=0$. On the  other hand the energy of the
resulting configuration is $E=N_B - N_C$ since any exchange with a $B$
particle yields a cost of one unit while an exchange with a $C$ particle
yields a gain of one unit of energy. Therefore for self consistency
the two densities $N_B$ and $N_C$ have to be equal, and similarly,
they have to be equal to $N_A$.

The Hamiltonian (\ref{Hamiltonian}) may be used to calculate steady
state averages corresponding to the dynamics (\ref{eq:dynamics}).
We start by an outline of the calculation of the free energy.
Consider a ground state of the system (\ref{Groundstate}). The low
lying excitations around this ground state are obtained by exchanging nn
pairs of particles around each of the three domain walls.
Let us first examine excitations which are localized around one of
the walls, say, $AB$. An
excitation can be formed by one or more $B$ particles moving into the
$A$ domain (equivalently $A$ particles moving into the $B$ domain). A
moving $B$ particle may be considered as a walker. The energy of the
system increases linearly with the distance traveled by the walker
inside the $A$ domain. An excitation of energy $m$ at the $AB$
boundary is formed by $j$ walkers passing a total distance of
$m$. Hence, the total number of states of energy $m$ at the $AB$
boundary is equal to the number of ways $P(m)$ of partitioning an
integer\index{partioning an integer} $m$ into a sum of (positive) integers. This and related
functions have been extensively studied in the mathematical literature
over many years. Although no explicit general formula for $P(m)$ is
available, its asymptotic form for large $m$ is known \cite{Partition}
\begin{equation}
P(m) \simeq \frac{1}{4m \sqrt{3}} \exp{(\pi (2/3)^{1/2} \ m^{1/2})} .
\end{equation}
Also, a well known result attributed to Euler yields the generating
function
\begin{equation}
\label{Y}
Y=\sum_{m=0}^{\infty} q^m P(m) = \frac{1}{(q)_{\infty}}  ,
\end{equation}
where
\begin{equation}
(q)_{\infty}=\lim_{n\rightarrow \infty} (1-q)(1-q^2)\ldots (1-q^n)  .
\end{equation}
This result may be extended to obtain the partition sum $Z_N$ of the 
full model. In the limit of large $N$ the three domain walls basically
do not interact. It has been shown that excitations around the different
domain boundaries contribute additively to the energy spectrum \cite{ABC}.
As a result in the thermodynamic limit the partition sum takes the form
\begin{equation}
\label{Z}
Z_N = N/[(q)_{\infty}]^3  , 
\end{equation} 
where the multiplicative factor $N$ results from the $N-$fold degeneracy
of the ground state and the cubic power is related to the three 
independent excitation spectra associated with the three domain walls.

It is of interest to note that the partition sum is linear and not
exponential in $N$, as is usually expected, meaning that the free
energy is not extensive. This is a result of the long-range interaction
in the Hamiltonian and the fact that the energy excitations are localized
near the domain boundaries.

Whether or not a system has long-range order in the steady state can
be found by studying the decay of two-point density correlation
functions. For example the probability of finding an $A$ particle at
site $i$ and a $B$ particle at site $j$ is,
\begin{equation}
\label{CorrF1}
\langle A_i B_j \rangle  = \frac{1}{Z_N} \sum_{\{X_k\}}
A_i B_j
 ~q^{{\cal H}(\{X_k\})}  ,
\end{equation}
where the summation is over all configurations $\{X_k\}$ in which
$N_A=N_B=N_C$. Due to symmetry many of the correlation functions will
be the same, for example $\langle A_i A_j \rangle = \langle B_i B_j
\rangle =\langle C_i C_j \rangle$. A sufficient condition for
the existence of phase separation is

\begin{equation}
\label{CorrF2}
\lim_{r\rightarrow \infty}~\lim_{N\rightarrow \infty} (\langle A_1 A_r
\rangle - \langle A_1 \rangle \langle A_r \rangle) >0  .
\end{equation}
Since $\langle A_i \rangle =1/3$ we wish to show that
$\lim_{r\rightarrow \infty} ~\lim_{N\rightarrow \infty} \langle A_1
A_r \rangle > 1/9$.  In fact it can be shown \cite{ABC}
that for any given $r$ and for sufficiently large $N$,
\begin{equation}
\label{CorrF3}
\langle A_1 A_r \rangle = 1/3 - {\cal O}(r/N)  .
\end{equation}
This result not only demonstrates that there is phase separation, but
also that each of the domains is pure. Namely the probability of
finding a particle a large distance inside a domain of particles of
another type is vanishingly small in the thermodynamic limit.

Numerical simulations of the model for the case of unequal densities,
where such analysis cannot be carried out, strongly indicate
that phase separation takes place as long as none of the three densities
vanish. They also indicate that the coarsening process which accompanies
phase separation is rather slow, with the characteristic length
diverging like  $\ln t$  at long times.

\section{S\lowercase{ummary}}
\label{Summary}

In these lecture notes some collective phenomena which occur
in one-dimensional driven systems have been reviewed.
These systems have been extensively studied in recent
years by introducing simple models and analysing their
steady state properties. Some of these models have been
demonstrated to exhibit a rich variety of phenomena which are
unexpected in equilibrium one-dimensional systems. 

Simple asymmetric exclusion processes in open systems
were shown to exhibit both first order and continuous phase
transitions. Other systems which have in the past been 
demonstrated to exhibit phase transitions in $1d$
are directed percolation  \cite{DK,Kinzel} and contact processes
\cite{Durrett}. These system, however, possess one or more
{\it absorbing states}. Once the system
evolves into one of these states the dynamics is such that the
system is unable to exit. Under these conditions, the existence of
a phase transition between a trapped and an untrapped states  is rather
natural. Usually, once the dynamics in these models is generalised
to allow for an exit from the absorbing state no phase transition
takes place. The phase transitions occurring in the
asymmetric exclusion processes discussed in this paper
are rather different, as the dynamics in these models does not possess
absorbing states.

Mechanisms which lead to spontaneous symmetry breaking and phase 
separation in one-dimensional nonequilibrium systems have been
discussed. A common crucial feature of these models is
that the dynamics conserves, at least to some degree, the order
parameter. In the `bridge' model, the densities of the two
types of particles are conserved in the bulk although they
are not conserved at the two ends of the system. In the $ABC$
model, on the other hand, the three densities are fully conserved.
When non-conserving processes are introduced into these models
spontaneous symmetry breaking and phase separation do not take place.
It would be very interesting to consider the possibility of
spontaneous symmetry breaking in one dimension when the dynamics does not
conserve the order parameter. 
A related problem has been considered in the context or error correcting
computation algorithems. An example of a one-dimensional
array of coupled probabilistic cellular automata has been constructed
and shown to yield breaking of ergodicity, as would a model
with spontaneous symmetry breaking \cite{Gacs}. This approach suggests
that indeed spontaneous symmetry breaking in $1d$ may exist even
when the dynamics is not conserving. However the example given
is rather complicated and not well understood. 

In spite of the progress made in recent years in the understanding of
nonequilibrium collective phenomena, many basic questions remain open,
even for the restricted and relatively simple class of systems which evolve
into a steady state.  For example a classification of continuous nonequilibrium
transitions into universality classes, like the one which exists for equilibrium
transitions, is not available. Also the dynamical process of the approach to steady
state is far from being understood in many cases. The approach outlined in these
notes, which involves constructing simple dynamical models and analysing the resulting
collective behaviour may prove to be helpful in developing better understanding of
some of these complex questions.

\vspace{0.5cm}
\noindent {\bf Acknowledgments}

I thank Martin Evans and Yariv Kafri for many useful comments and for
critical reading of the manuscript.

\end{document}